\documentstyle[12pt,epsfig,here,twoside]{article}

\begin{document}
\hspace*{10cm}
\normalsize{IHEP\,--\,HD/96\,--\,6}
\begin{center}
{\LARGE \bf Baryon Production \\[2mm]
            in ALEPH}
\\[5mm]
Presented on behalf of the ALEPH Collaboration by
\\[5mm]
{\bf U.\ Becker}
\footnote{Talk given at the HADRONS 96 conference 9.6.\,--\,16.6.\ 1996 in NOVY SVET, Ukraine }
\\[3mm]
{ \it Institut f\"ur Hochenergiephysik der Universit\"at Heidelberg \\
Heidelberg, Germany}
\\[5mm]
{\large \bf Abstract}
\end{center}
\begin{quote}
Several recent results of the ALEPH Collaboration covering different aspects 
of baryon production on the Z resonance are presented. In particular 
production rates of hyperons, the full kinematical reconstruction of the 
$\Lambda_b$, 
observation of $\Xi_b$ in its semileptonic decay, and the measurements of the 
polarization of $\Lambda$ and $\Lambda_b$ baryons are discussed.
\end{quote}
\newpage
\section{Introduction}
Baryon
production in quark fragmentation is still poorly understood. 
The existing models reproduce the available
data quite well.  
To distinguish between these models and to come to a more fundamental understanding
of the production mechanism of baryons more experimental input is needed. Results of the 
ALEPH Collaboration on hyperon and b\,--\,baryon production in Z decays are reported here. 
Whenever  
possible, not only the overall production rates but also the momentum spectra 
of baryons have been investigated and compared to Monte Carlo predictions.
\par
While the number of produced hyperons is quite large  the production rate of 
b\,--\,baryons is low. 
Only few of them have been observed by LEP experiments so far. 
The full reconstruction and mass determination of the $\Lambda_b$ and the 
observation of the $\Xi_b$ through $\Xi$\,--\,lepton correlations will be described.
\par
Z decays provide many possibilities to test the 
Standard Model.    
Quarks produced in Z decays are expected to obtain a strong longitudinal 
polarization due to parity violation in the electroweak interaction. 
While mesons always cascade down to spin zero pseudoscalar states retaining
no polarization information, baryons might preserve a 
large fraction of the initial quark polarization. Recent measurements of the 
$\Lambda$ and $\Lambda_b$ polarizations are presented here.\par
The data sample used for these studies has been taken with the detector ALEPH 
at the LEP storage ring at a center of mass energy of about 91\,GeV.
The detector and its performance are described in detail elsewhere \cite{detek}.
\section{Hyperon production}
ALEPH has studied recently the inclusive production rates of several octet and 
decuplet hyperons: $\Sigma^{\pm}(1385)$, $\Xi^-$, $\Xi^0(1530)$ and 
$\Omega^-$\cite{hyper}. Except for the $\Omega^-$ also the energy spectra have been 
measured. The four hyperons have been reconstructed in their 
decays:\par
\begin{center}
\begin{tabular}{lcl}
$\Sigma^{\pm}(1385)$&$\rightarrow$&$\Lambda\pi^{\pm}$\\
$\Xi^-$&$\rightarrow$&$\Lambda\pi^-$\\
$\Xi^0(1530)$&$\rightarrow$&$\Xi^-\pi^+$\\
$\Omega^-$&$\rightarrow$&$\Lambda\mbox{K}^-$\\
\end{tabular}
\end{center}
where the $\Lambda$ candidates are tagged in the decay channel 
$\Lambda \rightarrow p \pi^-$. The resulting mass distributions are shown in 
Fig.\,1a\,-\,d.\\
Except for the $\Omega^-$ (due to low statistics) 
the energy spectra have been determined and compared with  
Monte Carlo predictions from JETSET 7.3\cite{lund} and UCLA\cite{ucla} (Fig.2). 
The agreement is  quite reasonable with slight preferences for JETSET.\\
The mean multiplicities per Z decay have then been determined using JETSET to extrapolate 
into the inaccessible low energy region. 
\begin{figure}[H]
\begin{center}
\vspace*{-3cm}
\epsfig{file=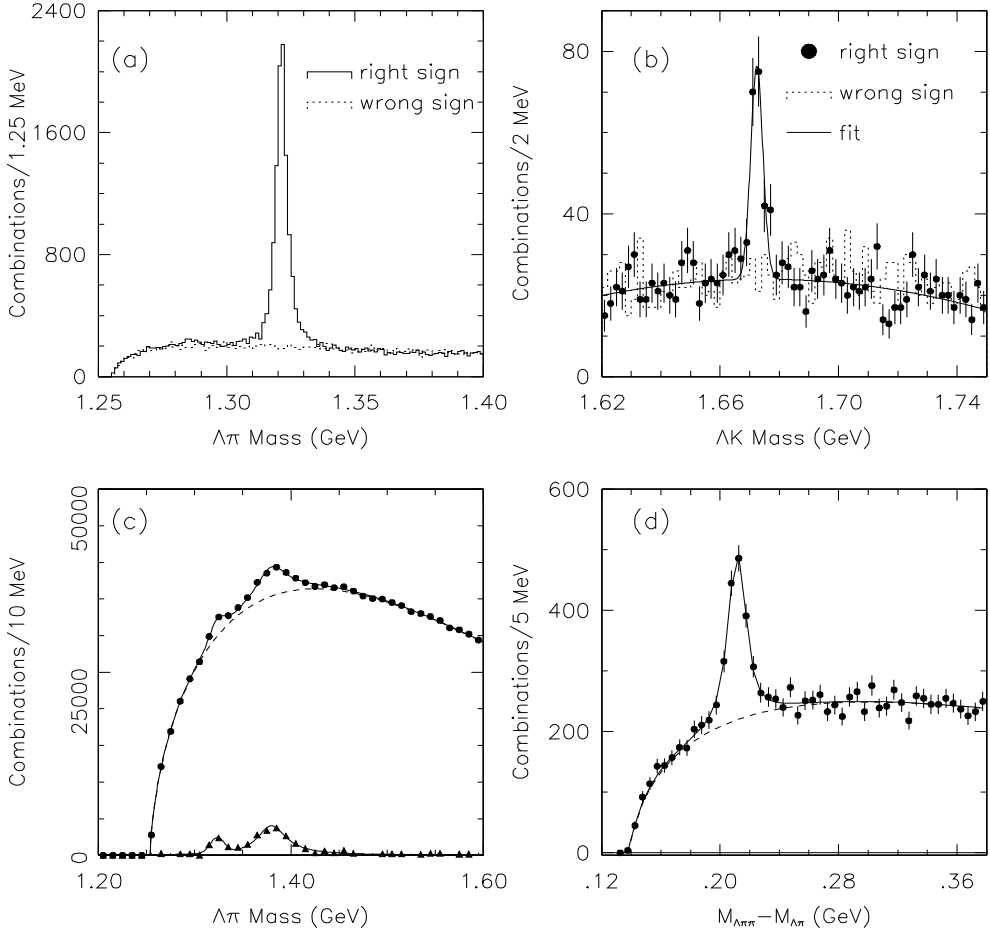,height=16.0cm,width=16.0cm}
\caption{(a)\em{The signal for $\Xi^-\rightarrow\Lambda\pi^-$. (b) The signal 
for $\Omega\rightarrow\Lambda\mbox{K}^-$. (c) The mass spectrum for 
$\Sigma^{\pm}(1385)\rightarrow\Lambda\pi^{\pm}$, fit to a background 
shape, a $\Xi^-$ contribution, plus a Breit\,--\,Wigner resonance. Also shown 
is the background\,--\,subtracted spectrum. (d) The mass\,--\,difference 
spectrum for $\Xi^0(1530)\rightarrow\Xi^-\pi^+$ candidates, fit to the 
background shape plus a Breit\,--\,Wigner resonance convolved with a gaussian.}}
\end{center}
\end{figure}
The preliminary results are:
\begin{center}
\begin{tabular}{rcl}
$\langle N_{\Xi^-}\rangle + \langle N_{\bar{\Xi}^+}\rangle$& = &$0.0285\pm0.0007\pm0.0020$\\
$\langle N_{\Xi^0(1530)}\rangle + \langle N_{\bar{\Xi}^0(1530)^0}\rangle$& = &$0.0072\pm0.0004\pm0.0006$\\
$\langle N_{\Sigma^{\pm}(1385)}\rangle + \langle N_{\bar{\Sigma}(1385)^\pm}\rangle$& = &$0.065\pm0.004\pm0.009$\\
$\langle N_{\Omega^-}\rangle + \langle N_{\bar{\Omega}^+}\rangle$& = &$0.0010\pm0.0002\pm0.0002$\\
\end{tabular}
\end{center}
While our measurement of the $\Sigma^{\pm}(1385)$ rate is in good agreement with the 
Monte Carlo prediction it is nevertheless about 1.7 times higher than the corresponding 
measurements from DELPHI and OPAL\cite{opal_sig},\cite{del_sig},\cite{opal_sig2}. Fig.3 gives an overview about the hyperon production 
rates at LEP\cite{regul}.\newpage
\begin{figure}[H]
\begin{center}
\epsfig{file=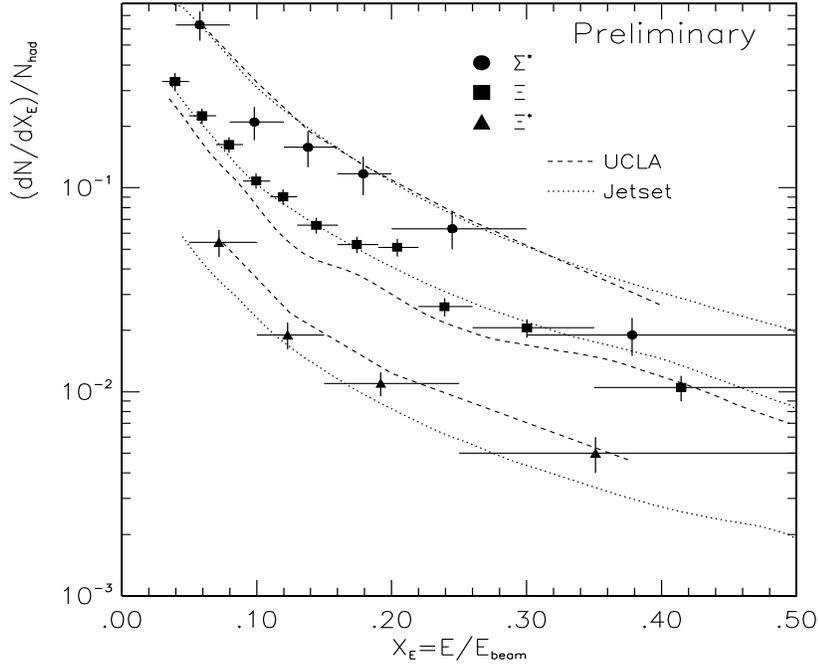,height=12.0cm,width=13.0cm}
\vspace* {-2.0cm}
\caption{\em{The measured $x_E$ distribution for $\Sigma^{\pm}(1385)$, $\Xi^-$, and
$\Xi^0(1530)$ (with antiparticles included), compared with predictions from the 
UCLA and JETSET 7.3  models.}}
\end{center}
\end{figure}
\vspace* {-2.0cm}
\begin{figure}[H]
\begin{center}
\epsfig{file=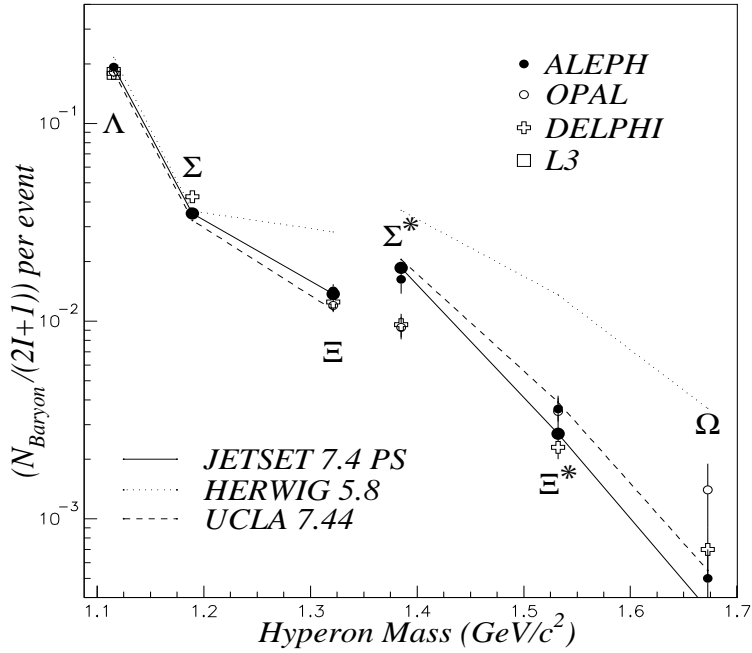,height=10.0cm,width=11.0cm}
\caption{\em{Production rates of hyperons at LEP together with predictions of JETSET, 
HERWIG and UCLA models. Lines are drawn to guide the eye.}}
\end{center}
\end{figure}
\newpage
\section{b\,--\,baryons}
The situation in the sector of heavy baryons containing a b\,--\,quark is quite 
different.  
Due to their low production rates combined with 
small branching ratios (in the order of a few $\%$) and long decay chains 
only very few  b\,--\,baryons have been observed at LEP so far, namely the $\Lambda_b$, $\Xi_b$ 
(which will be presented here) and $\Sigma_b$, $\Sigma_b^{\ast}$ by the DELPHI Collaboration \cite{sigb}. 
\subsection{The $\Lambda_b$}
The $\Lambda_b$ has been observed indirectly since several years in  $\Lambda l^-$ and $\Lambda_c^+ l^-$ 
correlations (where $l$ = muon or electron). Its production rate has been determined by e.g.\ ALEPH to be $Br(b\rightarrow\Lambda_b)\cdot
Br(\Lambda_b\rightarrow\Lambda_c^+l\bar{\nu}X)=(0.76\pm0.15\pm0.13)10^{-3}$\cite{lbprod} but its mass remained only poorly known. Now the $\Lambda_b$ has 
been fully 
reconstructed in its decay channel $\Lambda_b\rightarrow\Lambda_c^+\pi^-$ with 
$\Lambda_c^+\rightarrow pK\pi$, $p\bar{K}^0$ or $\Lambda\pi\pi\pi$. Several cuts have been 
applied to reduce the background.
The result is shown in Fig.4\\
The background has been estimated to be $0.38\pm0.05$ entries leaving us with a statistical 
significance of the observed peak at the 3.3\,$\sigma$ level, not taking 
into account the mass clustering of the
four events.
The $\Lambda_b$ mass has been found to be $(5614\pm21(stat.)\pm4(sys.))\mbox{MeV/c}^2$\cite{lambal}. This 
is in good agreement with the recent measurements of DELPHI $(5668\pm16\pm8)\mbox{MeV/c}^2$\cite{lambdel} 
and CDF $(5623\pm5\pm4)\mbox{MeV/c}^2$\cite{lambcdf} and the PDG94 value $(5641\pm50)\mbox{MeV/c}^2$\cite{pdg}.
\begin{figure}[H]
\begin{center}
\unitlength1cm
\vspace*{-1.0cm}
\begin{minipage}[b]{8.0cm}
\begin{picture}(8.5,10.0)
\epsfig{file=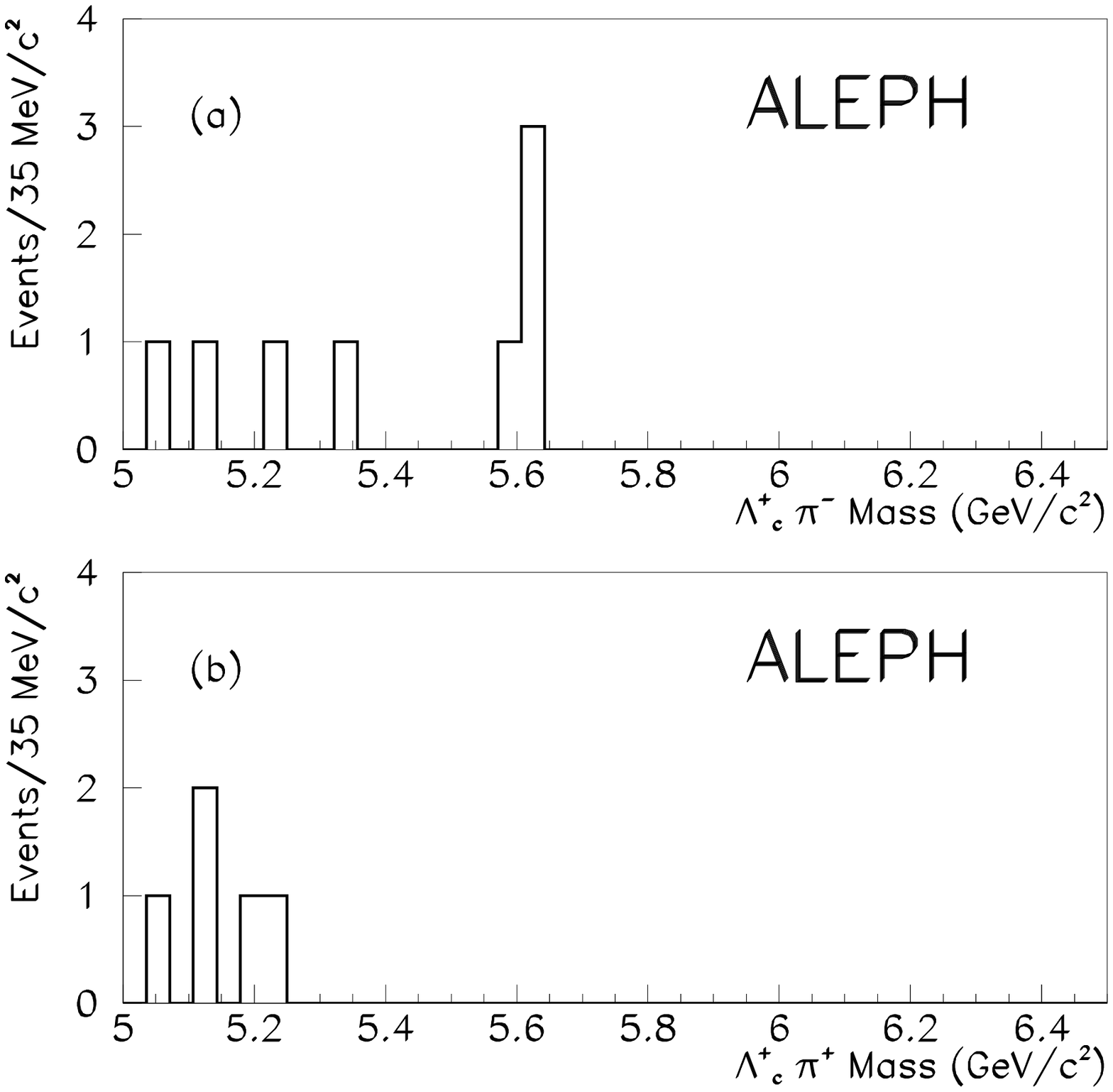,height=9.5cm,width=8.5cm}
\end{picture}
\end{minipage}
\hfill
\begin{minipage}[b]{8.0cm}
\begin{picture}(8.5,10.0)
\epsfig{file=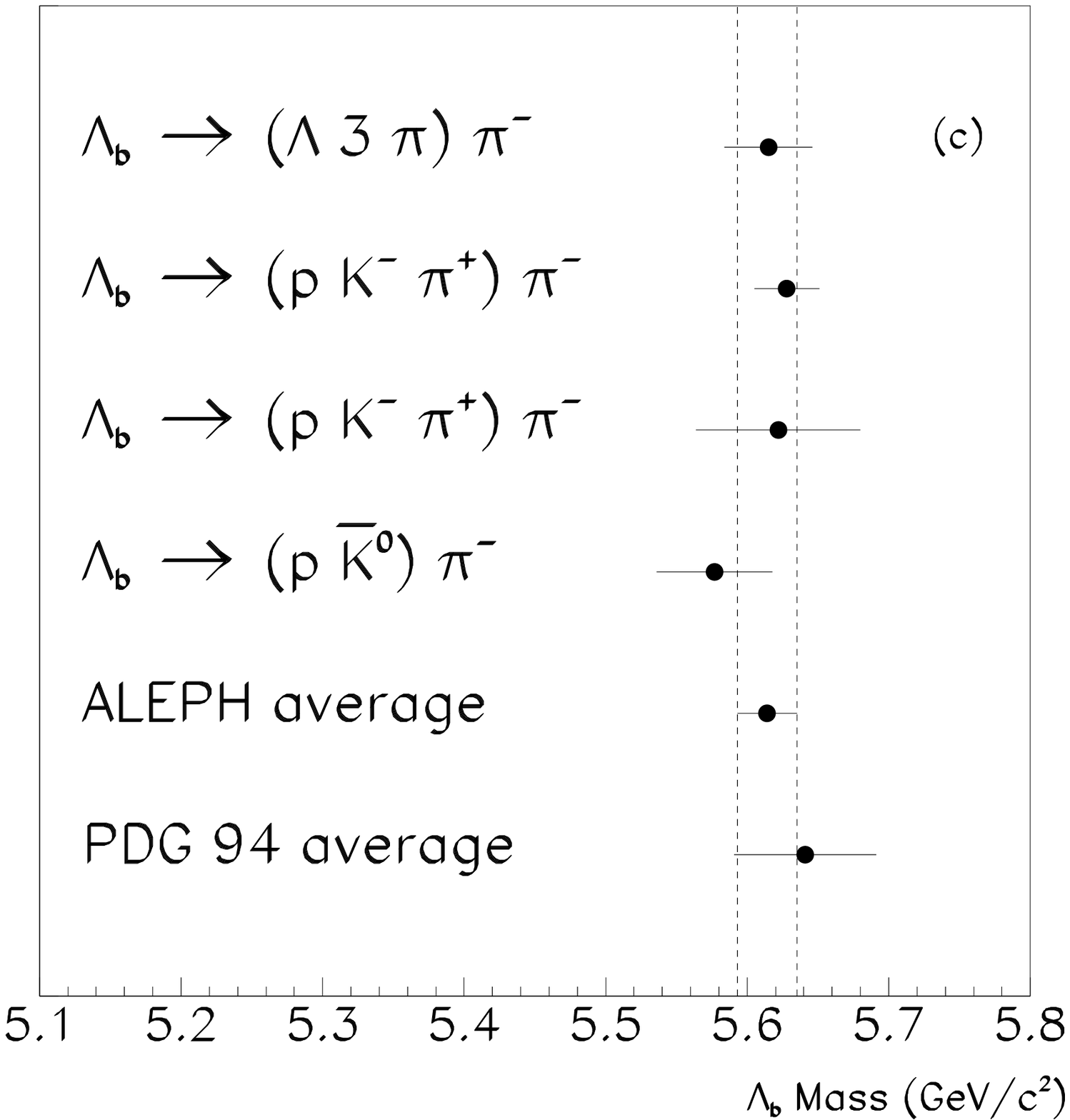,height=9.5cm,width=8.5cm}
\end{picture}
\end{minipage}
\caption{\em{(a) $\Lambda_c\pi$ invariant mass distributions for the right\,--\,sign combination and 
(b) for the wrong\,--\,sign combinations. (c) $\Lambda_b$ invariant masses for the four selected candidates. 
The dotted lines indicate the $\pm1\sigma$ values around the ALEPH average measurement.}}
\end{center}
\end{figure}
\newpage
\subsection{The $\Xi_b$}
A search for $\Xi^-l^-$ correlations (again $l$ stands for muon or electron) in the same hemisphere of the detector has been made to 
find evidence for $\Xi_b$ production. After correcting for other
possible sources of $\Xi^-l^-$ pairs as e.g.\ $\Lambda_b$ B or accidentals
an excess of $22.5\pm5.7$ like\,--\,sign combinations with 
respect to opposite\,--\,sign combinations has been observed. It is therefore interpreted as 
from the semileptonic decay of the $\Xi_b$ and can 
be translated into the following product of branching fractions:
\begin{displaymath}
Br(b\rightarrow\Xi_b)\cdot
Br(\Xi_b\rightarrow\Xi^-l^-\bar{\nu}X)=(5.3\pm1.3\pm0.7)10^{-4}
\end{displaymath}
per lepton species\cite{xibal}. This value is in good agreement with the DELPHI result\cite{xibdel}.
\section{Polarization}
Polarization of the primary quarks coming from Z decays is due to the parity violating  
inequality of the Z coupling to right and left handed 
fermions. Ignoring mass 
effects and gluon radiation (which may depolarize the quarks by about $3\%$ \cite{korn}) the expected 
polarization can be written as
\begin{displaymath}
P^q_L={\frac{-2g_vg_a}{g^2_V+g^2_A}}
\end{displaymath}
where $g_v$ and $g_a$ are the vector and axial vector coupling constants of the Z to the quark.
With $\sin^2\vartheta_W$ = 0.23 a polarization of $-94\%$ for downtype 
quarks is expected according to the Standard Model, but may be diluted substantially during the 
hadronization phase due to spin\,--\,spin forces.\\
Mesons always cascade down to spin zero pseudoscalar particles which do not retain 
any polarization information. In contrast baryons are expected to 
preserve a sizeable fraction of the initial quark polarization which, however, gets
smaller 
if the weakly decaying ground state baryon is produced in the decay of a higher 
resonance\cite{close},\cite{falk}. Therefore the mean measureable polarization will depend on the production rate of 
these resonances and will thus give additional information about the hadronization phase.

\subsection{The $\Lambda$ polarization}
The polarization of $\Lambda$ hyperons can be measured  by investigating the decay angle 
$\vartheta^{\ast}$ in their decay $\Lambda \rightarrow p \pi^-$. Here $\vartheta^{\ast}$ 
is defined as the angle between the proton momentum in the $\Lambda$ restframe and
the $\Lambda$ boost direction.
The $\vartheta^{\ast}$ distribution is expected to follow $R(\vartheta^{\ast}) = 1 + \alpha
P_L^{\Lambda}\cos \vartheta^{\ast}$ with the asymmetry parameter $\alpha = 0.642\pm0.013$ \cite{pdg} and 
the $\Lambda$ polarization $P_L^{\Lambda}$. 
The observed $\Lambda$ polarization is diluted with respect to the primary s\,--\,quark 
polarization by those $\Lambda$ baryons which do not contain a primary quark. To derive a Monte Carlo 
prediction of $P_L^{\Lambda}$, first the fraction of $\Lambda$'s containing a primary quark 
has been determined in the data by measuring the $\Lambda$ forward\,--\,backward asymmetry and the 
correlation of back\,--\,to\,--\,back $\Lambda\bar{\Lambda}$ pairs. 
The differences between data and MC have been used as 
correction factors for all possible sources of polarized $\Lambda$'s (see \cite{lpolal} for details). Then the fraction of 
$\Lambda$ baryons coming from e.g.\ $\Lambda_b,\Sigma$ etc.\ has been corrected, and the 
expected polarizations \cite{hakk} have been added up. The ALEPH  
measurement is shown in Fig.5 together with the expectation. For $\Lambda$'s with $z=p(\Lambda)/P_{Beam} > 0.3$ the 
polarization has been determined to be $P_L^{\Lambda} = -0.32\pm0.07$ \cite{lpolal}, in good agreement 
with the JETSET prediction of $-0.39\pm0.08$.
\begin{figure}[H]
\begin{center}
\unitlength1cm
\vspace*{-0.5cm}
\begin{minipage}[b]{8.0cm}
\begin{picture}(8.5,10.0)
\epsfig{file=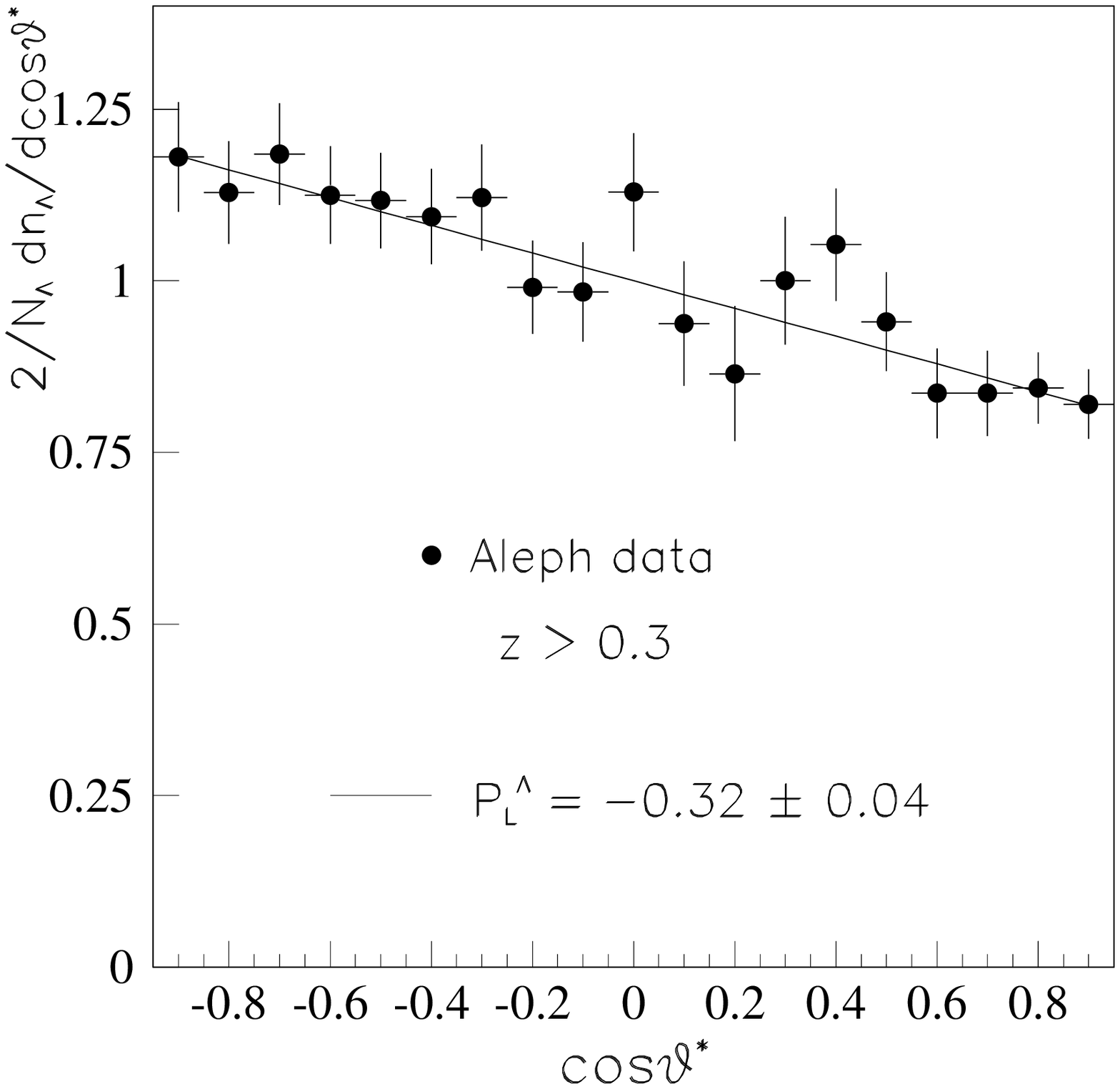,height=10.0cm,width=8.5cm}
\end{picture}
\end{minipage}
\begin{minipage}[b]{8.0cm}
\begin{picture}(8.5,10.0)
\epsfig{file=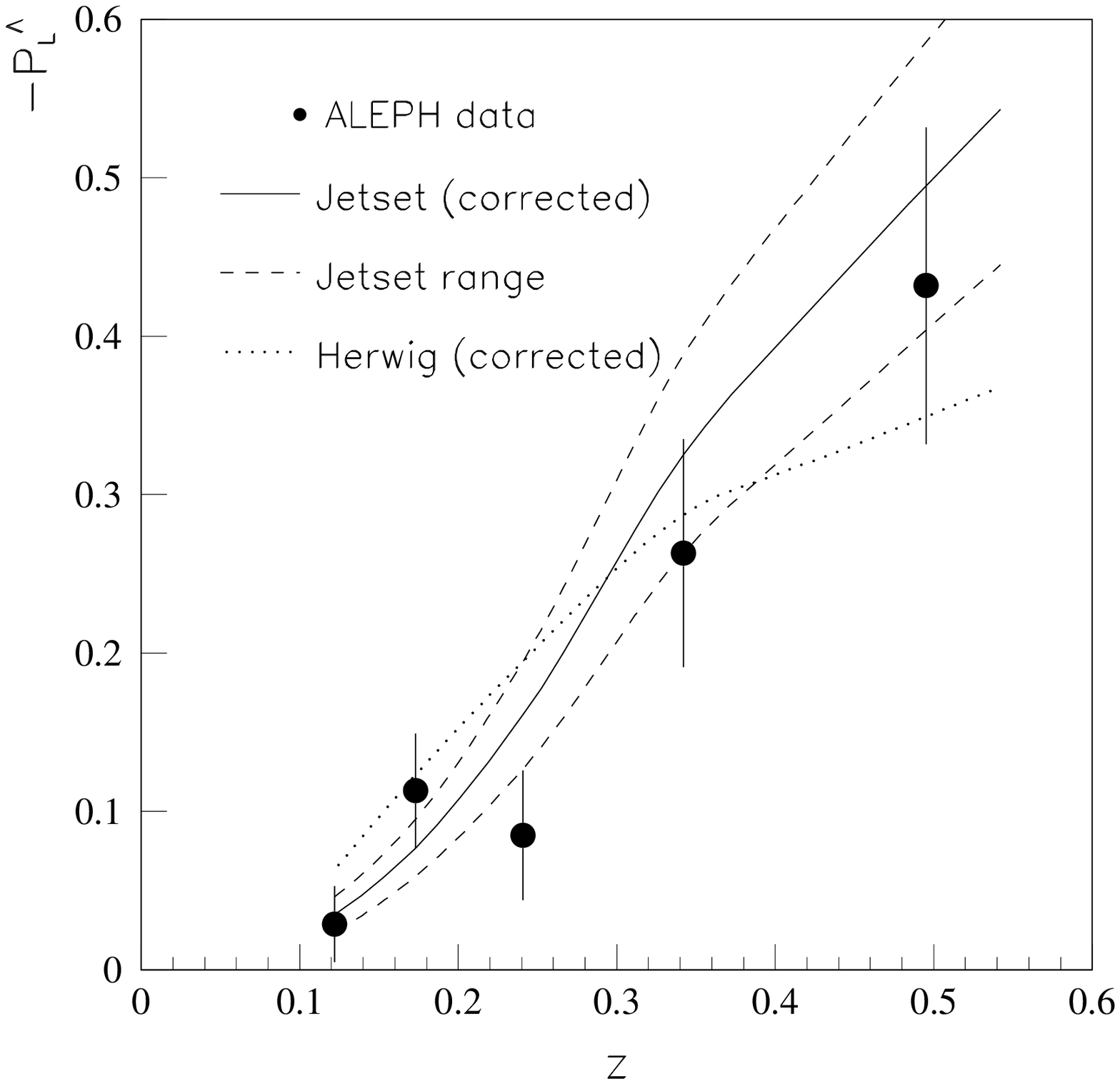,height=10.0cm,width=8.5cm}
\end{picture}
\end{minipage}
\hfill
\vspace*{-0.5cm}
\caption{\em{{\bf Right}: Fit of the longitudinal $\Lambda$ polarization to the decay angle 
distribution for $z>0.3$ .}\newline
{\bf Left}: The measured longitudinal $\Lambda$ polarization is shown as as function of z together with the 
JETSET 
prediction after multiplying by the correction factor (shown as solid line). The dashed lines indicate 
the estimated uncertainty of the JETSET prediction. The HERWIG prediction after multiplication by the 
correction factor is shown as dotted line.
}
\end{center}
\end{figure}
\subsection{The $\Lambda_b$ polarization}
There is no need to estimate the `primary' $\Lambda_b$ fraction since b\,--\,quarks   
nearly always originate directly from Z decays. It has been suggested for some time to measure the 
$\Lambda_b$ polarization (which is equal to the b\,--\,quark polarization in the heavy quark 
limit) in its semileptonic decay\cite{lpolsl}.\\
Apart of the interest in the polarization itself,  
weak decays of polarized b\,--\,quarks provide another interesting aspect. It has been pointed out 
by Gronau and Wakaizumi \cite{gronau} that the left handedness of the b\,--\,quark coupling to
the charged weak current has never been proven. They have shown that, assuming the same 
handedness for the leptons involved in the decay, the chirality could not be determined in 
b\,--\,meson decay . 
Amundsen et.\ al.\ \cite{amun} suggested to resolve this ambiguity through an investigation of the 
semileptonic decay of {\it polarized} b\,--\,quarks which are available in $\Lambda_b$'s from the 
Z resonance. When deriving the polarization from the lepton spectra, a V+A coupling 
would result in a change of sign of the measured with respect to the 
true polarization.
This is in principle analogous to the determination of the Michel parameter $\xi$ in  
$\tau$ decays.\par
The ALEPH analysis of the $\Lambda_b$ polarization \cite{lbpolal} has followed a suggestion of Bonvicini and 
Randall \cite{bonvi} who proposed to study the ratio $y$ of the mean energies of charged leptons and  
neutrinos from semileptonic $\Lambda_b$ decays. This procedure offers several advantages: 
Independence from the $\Lambda_b$ fragmentation 
function, greater sensitivity than the charged lepton alone and more statistics than any 
exclusive decay.\\
For this purpose $\Lambda\pi^+l^-$ pairs have been selected. The neutrino energy is 
calculated from the visible amount of energy in the detector: $E_{\nu} = E_{tot}-E_{vis}$, 
with $E_{tot} = \sqrt{s}/2 + (M_{same}^2- M_{oppo}^2)/2\sqrt{s}$. Here $M_{same}$ ($M_{oppo}$) 
is the invariant mass in the same (opposite) hemisphere. Biases due to selection cuts and neutrino energy 
reconstruction have been corrected with 
Monte Carlo simulations.
\vspace*{-1.5cm}
\begin{figure}[H]
\begin{center}
\epsfig{file=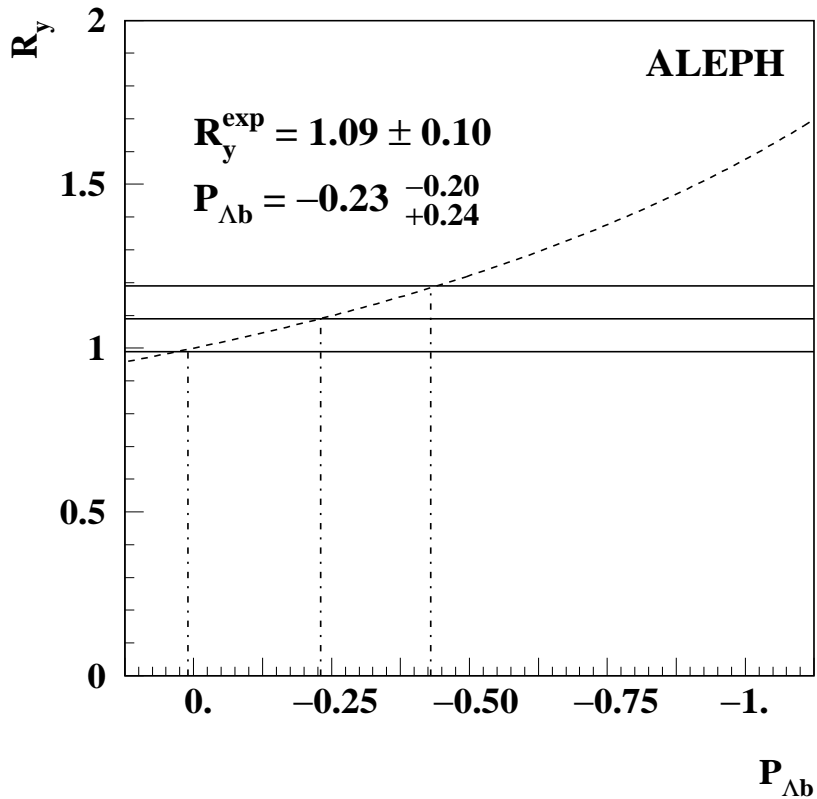,height=12.0cm,width=12.5cm}
\caption{\em{The method to extract the $\Lambda_b$ polarization value: Comparison between the measured 
$R^{exp}_y$ value and the theoretical prediction $R_y$}}
\end{center}
\end{figure}
The polarization in the data is extracted by comparing $R^{exp}_y=y_{data}/y_{MC}(P=0)$ (where $y_{MC}(P=0)$
is the expected ratio between lepton and neutrino mean energy without any polarization) 
with the expected $R_y$ curve and correcting for differences between Data and Monte Carlo (Fig.6):
\begin{displaymath}
{\cal P}^{\Lambda_b}=-0.23^{+0.24}_{-0.20}(stat.)^{+0.08}_{-0.07}(syst.)
\end{displaymath}
This result can be compared with an expectation of about -0.73 estimated from \cite{falk} 
together with the $\Sigma_b^{(\ast)}$ production rates measured by DELPHI \cite{sigb}. It is quite low (but 
consistent with the expectation), 
indicating 
additional depolarization effects during hadronization.

\end{document}